\documentclass[10pt]{iopart}

\usepackage{subfigure}
\usepackage{upgreek}
\usepackage{xcolor}
\usepackage{float}
\usepackage{hyperref}
\usepackage{mathtools}

\usepackage{graphicx}
\usepackage{dcolumn}
\usepackage{bm}

\usepackage[utf8]{inputenc}
\usepackage[T1]{fontenc}
\usepackage{mathptmx}
\usepackage{etoolbox}

\makeatletter
\def\@email#1#2{%
 \endgroup
 \patchcmd{\titleblock@produce}
  {\frontmatter@RRAPformat}
  {\frontmatter@RRAPformat{\produce@RRAP{*#1\href{mailto:#2}{#2}}}\frontmatter@RRAPformat}
  {}{}
}%
\makeatother
\begin{document}

\title{Phase classification using neural networks: application to supercooled, polymorphic core-softened mixtures}

\author{V F Hernandes$^1$, M S Marques$^2$ and
José Rafael Bordin$^3$}

\address{$^1$Programa de Pós-Graduação em Física, Departamento de Física, Instituto de Física e Matemática, Universidade Federal de Pelotas. Caixa Postal 354, 96001-970, Pelotas-RS, Brazil}
\address{$^2$ Centro das Ciências Exatas e das Tecnologias, Universidade Federal do Oeste da Bahia\\Rua Bertioga, 892, Morada Nobre, CEP 47810-059, Barreiras-BA, Brazil}
\address{$^3$ Departamento de Física, Instituto de Física e Matemática, Universidade Federal de Pelotas. Caixa Postal 354, 96001-970, Pelotas-RS, Brazil}
\ead{jrbordin@ufpel.edu.br}

\begin{indented}
\item[]August 2021
\end{indented}

\begin{abstract}
Characterization of phases of soft matter systems is a challenge faced in many physical chemical problems. For polymorphic fluids it is an even greater challenge. Specifically, glass forming fluids, as water, can have, besides solid polymorphism, more than one liquid and glassy phases, and even a liquid-liquid critical point. In this sense, we apply a neural network algorithm to analyze the phase behavior of a mixture of core-softened fluids that interact through the continuous-shouldered well (CSW) potential, which have liquid polymorphism and liquid-liquid critical points, similar to water. We also apply the neural network to mixtures of CSW fluids and core-softened alcohols models. We combine and expand methods based on bond-orientational order parameters to study mixtures, applied to mixtures of hardcore fluids and to supercooled water, to include longer range coordination shells. With this, the trained neural network was able to properly predict the crystalline solid phases, the fluid phases and the amorphous phase for the pure CSW and CSW-alcohols mixtures with high efficiency. More than this, information about the phase populations, obtained from the network approach, can help verify if the phase transition is continuous or discontinuous, and also to interpret how the metastable amorphous region spreads along the stable high density fluid phase. These findings help to understand the behavior of supercooled polymorphic fluids and extend the comprehension of how amphiphilic solutes affect the phases behavior.
\end{abstract}

%
%
%

%

%
\ioptwocol

\section{Introduction}

In the last decade, machine learning (ML) models successfully penetrated into virtually all areas of the scientific community, no longer being considered only an object of study \textit{per se}, but also a tool that can help solve the more diverse kind of problems faced by scientists \cite{carleo2019, buchanan2019, Schleder_2019, goh2017, mater2019}. Some few examples of ML applications in physics, chemistry and materials science include molecular and atomistic simulation \cite{noe20, hellstrom2019}, self-assembly of molecules \cite{Long2014, Reinhart2017, zhao2019, adorf2020}, force fields parametrization \cite{li2017, botu2017, mcdonagh2019}, soft-materials and proteins engineering \cite{ferguson2018, xu2020}, drug discovery \cite{klambauer19}. Another application that has recently being perfected and improved, taking advantage of ML models, is phase recognition. The task of identifying the structural formation of matter from local arrangements obtained from simulation data can be significantly refined with the utilization of statistical learning techniques. This approach is showing excellent results, such as local structure detection of polymorphic systems with supervised \cite{geiger2013} and unsupervised \cite{boattini2019} learning, feedforward neural network and recurrent neural network applied to local topological defects in confined liquid crystals detection~\cite{walters19}, identification of phases in matter \cite{carrasquilla17}, soft matter \cite{wei17, terao2020} and amorphous materials \cite{swanson2020} structures using convolutional neural networks and phase prediction of high-entropy alloys \cite{HUANG2019225, LEE2021109260}.

A system that is constantly under investigation, given its complexity, and that has been greatly benefited from a statistical approach, is water and its mixtures. Water is the solvent of life, and the main solvent in industry. Also, pure and ``simple" water presents more than 70 known anomalies \cite{chaplin2020url}, making it unique \cite{podgornik2011}. The origin of the high number of anomalies for temperatures in the supercooled regime can be related to a two liquids coexistence line that ends in a liquid-liquid critical point (LLCP), and to the competition between these liquids \cite{gallo2016, Bachler19, Pierre20}. At low densities, in the low density liquid phase, the water molecules have an ordered tetrahedral structure, while the high density liquid state is characterized by a more distorted tetrahedral structure, and with local higher density as consequence. The fact that water itself is a mixture of two liquids was hypothesized in the 90s, with an extensive theoretical debate since then, specially in the last decade \cite{poole1992, poole13, Limmer11, Limmer13, Palmer13a, Palmer18b, stanley1997, stanley1998, stanley2000, sciortino2003, debenedetti2003, handle2017}. Currently, the main theoretical evidences indicate that the LLCP exists, and the experimental evidences that support this conclusion are growing in the last years \cite{Amann13, Taschin13, Kim17, Caupin15, Hestand18, Kim978, salzmann2019}. However, it is an extremely hard task to achieve experimentally this region, known as ``no man's land", due to rapidly crystallization. In this sense, a computational approach is particularly useful. An extensive number of works have studied the region near the LLCP with molecular simulations using different potentials and approaches \cite{Palmer18b, handle2017, gallo2016, Debenedetti289}. More recently, ML algorithms are being utilized for recognizing the structures exhibited by water, specially nearby the second critical point. Distinct supervised learning approaches with Neural Networks are being explored, some of them based on bond-orientational order parameters \cite{martelli2020} or symmetry functions \cite{geiger2013}, others using data obtained from ab-initio calculations \cite{Gartner26040, monserrat2020} or even networks with more complex architectures, combining different methods \cite{fulford2019}. Other works studied the relation between structure and dynamics in the same supercooled region, but for general liquids and glasses, using unsupervised methods \cite{Schoenholz_2018, boattini2020}.

Along with research regarding pure water systems, a set of works have focused on the study of aqueous solutions in the supercooled regime. For instance, in the experimental work by Zhao and Angell \cite{Zhao16} the crystallization was repressed in the no man's land by adding ionic liquids that dissolve ideally in water, preventing the crystallization without destroying the water anomalies. Another class of solutions, which are simpler if compared to more complex systems, is the one of short-chain alcohols in water. It can be treated as a binary system, facilitating the computational approach. The motivation behind the studies of these particular systems lays on their wide range of application such as dispersion media \cite{champreda2012}, disinfectant \cite{smith1947}, in the food \cite{nguyen2020} and medical \cite{aspers2017} industries, among others. Many experimental and theoretical works have studied these short-chain alcohol/water mixtures \cite{PALINKAS199165, GONZALEZSALGADO2006161, corradini2012, munao2015, gonzalez-salgado2016, furlan2017, GONZALEZSALGADO2020112703}. In our recent works, we have performed Molecular Dynamics (MD) simulations with a core-softened (CS) approach to investigate the behavior of methanol-water \cite{MARQUES2020114420} and water mixtures with methanol, ethanol and 1-propanol \cite{marques2021}. In this approach, the waterlike solvent is modeled as the CSW fluid proposed by Franzese \cite{franzese2007}. Although it is a spherically core-softened potential with two length scales, without any directionallity and, therefore, is not water \cite{franzese2011}, this CS approach has been largely employed to understand water anomalies both in bulk and confined environments \cite{jagla1999, alan2008a,fomin11,bordin2016, bordin2018}. Particularly, the CSW model that we use in our work is able to reproduce the anomalous behavior of water in the supercooled regime, including the existence of two liquid phases whose coexistence line ends in the LLCP. Based on this potential, the alcohols are modeled as rigid polymers, as proposed by Urbic and co-authors \cite{urbic2015jcp}. Moreover, the hydroxyl group is modeled as a CSW bead, while the hydrophobicity of the polar sites is given by a modified Lennard Jones (LJ) potential. It was found \cite{marques2021} that the addition of distinct concentrations of alcohols with distinct chain lengths lead to the suppression of the crystal phase, with the favoring of the amorphous phase and the existence of the liquid-liquid phase transition - in addition to the waterlike anomalies in the supercooled regime. 
With a particular interest for this present work,
we observe a variety of phases: a high-density liquid (HDL) and a low-density liquid phase (LDL), two solid phases: a body-centered cubic one (BCC - phase I) and a hexagonal close-packed phase (HCP - phase II), and an amorphous solid phase (phase III). This polymorphism makes these mixtures great candidates to test if a NN algorithm is able to recognize the distinct phases in water-solvent mixtures in the supercooled regime.

Additionally, contrasting with the case of pure-water \cite{fulford2019, martelli2020} there is a lack of works applying ML models to autonomously recognize the phases of water-alcohol mixtures near the LLCP, an approach that can help to better understand the structural behavior of these systems. Given that, in this work, we set up a Neural Network based on Steinhardt parameters \cite{steinhardt1983, lechner2008}, adapted for binary mixtures, similar to what has been done by Boattini \textit{et al.} \cite{boattini2018}, capable of identify the phases of (methanol/ethanol/1-propanol)-water mixtures for different alcohol concentrations. Our goal is to check if this NN based approach can properly predict the phases and phase transitions and provide new insights about the polymorphism in the supercooled regime.

The paper is structured as follows. In Section II we describe the NN architecture and the parameters utilized for map the molecules' local structures. The results obtained, namely the phase diagram predicted by the NN for the different mixtures and concentrations, alongside a population analysis (the number of particles in each phase for a specific pressure-temperature configuration) are presented in Section III. In Section IV, we present a closing discussion, with the principal remarks and some perspectives for new works.

\section{The Computational Details}

The NN approach requires only a system snapshot. The last configuration from each $(N,P,T)$ simulation performed in our last work \cite{marques2021} was chosen as input for the NN. All the systems, for different alcohols and concentrations and pure water, are composed by $N = 1000$ molecules. Water and alcohol's hydroxyl groups are modeled as CSW particles, while the hydrophobic carbon chain in alcohols molecules are LJ sites. The waterlike solvent is monomeric, while the alcohols are linear rigid polymers: methanol is modeled as a dumbbell -- one CSW site, one LJ site, ethanol as a trimer -- one CSW site, two LJ sites -- and propanol as a tetramer -- one CSW site, three LJ sites. Detailed information about the simulation methods, parameters and the models can be found in our previous work \cite{marques2021}.
For the pure solvent and all mixtures cases, we have observed five distinct phases. Two liquid phases, namely Low Density Liquid (LDL) and High Density Liquid (LDL), separated by a first order coexistence line that ends in the LLCP. Also, above the LLCP, the Widom Line (WL) delimits the border between the LDL-predominant and HDL-predominant regions. Two crystalline phases were observed. At lower pressures, the system is in a BCC phase, that will be called of solid I, or just I for simplification, and in a HCP -- or solid phase II -- at intermediate pressures. Finally, at higher pressures, the system has an amorphous solid phase, or phase III. We called it an amorphous solid once it is disordered, with a structure similar to the HDL phase, but with no diffusion. The amorphous-HDL transition was characterized by an increase in the diffusion constant and by maxima in the isobaric expansion coefficient and in the specific heat at constant pressure \cite{marques2021}. All  quantities  with  an  asterisk  are  in  reduced  dimensionless units \cite{allen2017}.

To map the local environment of each particle of the system into a vector used as input for the ML model, a series of Bond-Order Parameters (BOOP) \cite{steinhardt1983, lechner2008} are calculated. The last configuration from the $(N,P,T)$ simulations are the input for the freud analysis python package \cite{freud2020} to calculate the BOOPs of the 1000 CSW particles in the system. The Voronoi tessellation was used to define nearest neighbors. For each particle $i$ with $N_b(i)$ neighbors, first we calculate $q_{lm} (i)$, $\bar{q}_{lm} (i)$ and $\bar{\bar{q}}_{lm} (i)$, defined as

\numparts
\begin{eqnarray}
 q_{lm} (i) = \frac{1}{N_b (i)} \sum_{j=1}^{N_b (i)} Y_{lm} (\Vec{r}_{ij}), \label{BOOP1a} \\
    \bar{q}_{lm} (i) = \frac{1}{N_b(i) + 1} \sum_{k \in \{i, N_b (i)\}} q_{lm} (k),\label{BOOP1b}\\
\bar{\bar{q}}_{lm} (i) = \frac{1}{N_b (i) + 1} \sum_{k \in \{i, N_b (i)\}} \bar{q}_{lm} (k), \label{BOOP1c}
\end{eqnarray}
\endnumparts

\noindent where $Y_{lm} (\Vec{r}_{ij})$ are the spherical harmonics for the distance vector $\Vec{r}_{ij}$ separating particle $i$ from $j$. Then, from (\ref{BOOP1a}), (\ref{BOOP1b}) and (\ref{BOOP1c}) we have  $q_{l} (i)$, $\bar{q}_{l} (i)$ and $\bar{\bar{q}}_{l} (i)$, respectively, given by

\begin{figure}[h!]
\begin{tabular}{c}
\hspace*{-0.4cm}                                                           

  \includegraphics[width=0.45\textwidth]{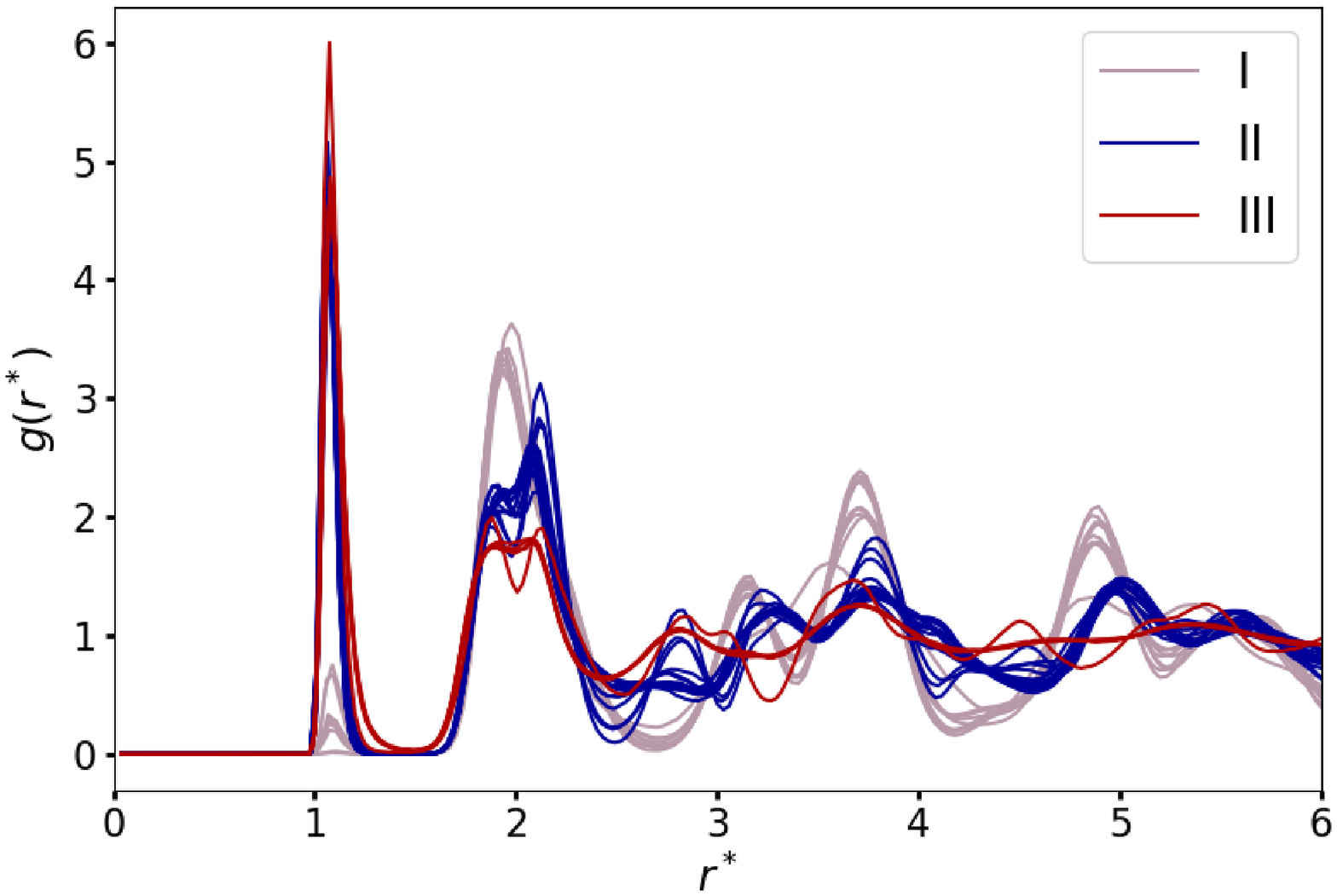} \\
  (a) \\ \includegraphics[width=0.45\textwidth]{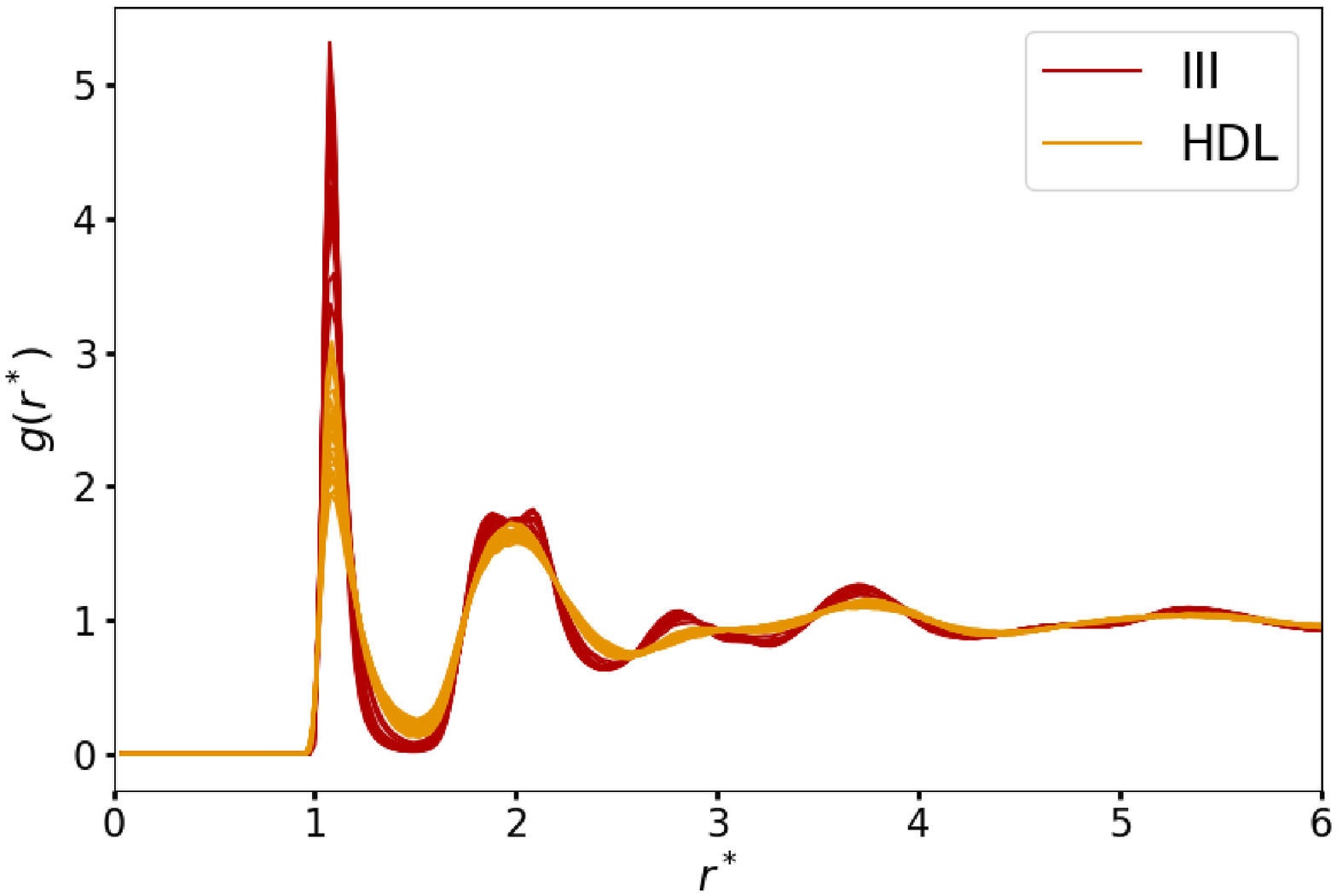} \\
(b)  \\
\end{tabular}
\caption{(a) Radial distribution function for pure water with $T^* = 0.26$. Grey lines are the pressures in the BCC phase, blue lines in the HCP phase and red in the amorphous phase. (b) Radial distribution function for pure water with $P^* = 0.28$. Red lines are the temperatures in the amorphous phase and yellow lines in the HDL phase.}
    \label{fig:g_de_r}
    \end{figure}

\numparts
\begin{eqnarray}
 q_{l} (i) = \sqrt{\frac{4 \pi}{2l + 1} \sum_{m=-l}^{l} \vert q_{lm} (i) \vert ^2 }, \label{BOOP2a} \\
    \bar{q}_{l} (i) = \sqrt{\frac{4 \pi}{2l + 1} \sum_{m=-l}^{l} \vert \bar{q}_{lm} (i) \vert ^2 },\label{BOOP2b}\\
\bar{\bar{q}}_{l} (i) = \sqrt{\frac{4 \pi}{2l + 1} \sum_{m=-l}^{l} \vert \bar{\bar{q}}_{lm} (i) \vert ^2 },\label{BOOP2c}
\end{eqnarray}
\endnumparts

\noindent where $q_{l} (i)$ is the rotational invariant BOOP, $\bar{q}_{l} (i)$ its average and $\bar{\bar{q}}_{l} (i)$ the average of the averages. The average values make it possible to get information about, approximately, the second-shell neighbors. Here, we introduced the average-average parameter, $\bar{\bar{q}}_{l} (i)$, which holds information regarding the structure of, approximately, third-shell neighbors. We want to test if this further parameter can be useful to uniquely identify a local structure that, for different configurations within the same phase, shows significant differences in the long-range coordination shells, as can be seen by analyzing the radial distribution function $g(r^*)$ for pure water with $T^* = 0.26$ at distinct pressures, in Figure \ref{fig:g_de_r}(a) and at distinct temperatures with fixed pressure equals to $P^* = 0.28$, in Figure \ref{fig:g_de_r}(b). The former case is that of a fixed temperature at which the three solid phases occur. The latter is the case of a fixed pressure at which the system presents an amorphous-HDL transition. 
 
\noindent Then, the cubic BOOPs, $w_{l} (i)$, their average, $\bar{w}_{l} (i)$, and the average of the average, $\bar{\bar{w}}_{l} (i)$,

\numparts
\begin{eqnarray}
 w_{l} (i) = \frac{\sum\limits_{m_1 + m_2 + m_3 = 0} \begin{psmallmatrix}
l & l & l\\
m_1 & m_2 & m_3
\end{psmallmatrix} q_{l m_1} (i) q_{l m_2} (i) q_{l m_3} (i)}{\big(\sum_{m=-l}^{l} \vert q_{l m} (i) \vert ^2 \big)^{3/2}}, \label{BOOP3a} \\[20pt]
    \bar{w}_{l} (i) =  \frac{\sum\limits_{m_1 + m_2 + m_3 = 0} \begin{psmallmatrix}
l & l & l\\
m_1 & m_2 & m_3
\end{psmallmatrix} \bar{q}_{l m_1} (i) \bar{q}_{l m_2} (i) \bar{q}_{l m_3} (i)}{\big(\sum_{m=-l}^{l} \vert \bar{q}_{l m} (i) \vert ^2 \big)^{3/2}},\label{BOOP3b}\\[20pt]
\bar{\bar{w}}_{l} (i) = \frac{\sum\limits_{m_1 + m_2 + m_3 = 0} \begin{psmallmatrix}
l & l & l\\
m_1 & m_2 & m_3
\end{psmallmatrix} \bar{\bar{q}}_{l m_1} (i) \bar{\bar{q}}_{l m_2} (i) \bar{\bar{q}}_{l m_3} (i)}{\big(\sum_{m=-l}^{l} \vert \bar{\bar{q}}_{l m} (i) \vert ^2 \big)^{3/2}},\label{BOOP3c} \\[15pt] \nonumber
\end{eqnarray}
\endnumparts

\noindent are calculated. Here, the term in parentheses corresponds to the Wigner $3j$ symbol. The set of parameters described so far should give a vector, composed by translationally and rotationally invariants, capable to uniquely describe the different phases' local environment. However, since we are dealing with binary-mixtures systems, a distinction between the parameters related to water molecules and those related to alcohol molecules is needed, specifically for the hydroxyl group of the alcohols' molecules, since they are modeled by the same potential interaction and parameters.
This common factor makes this approach for binary mixtures expandable to other alcohols, as long as the hydroxyl group is explicit in the simulation, since the bonds are always calculated considering the water-hydroxyl interaction.
Given that, we consider three variations for each parameter calculated. For instance, instead of a unique $q_l (i)$, we have $q_l^{W,A/A-W}(i)$, $q_l^{W,W}(i)$ and $q_l^{A,A}(i)$. The first term is the $q_l(i)$ parameter for $i$ being a water (alcohol) molecule considering all type of neighbors, water (alcohol) or alcohol (water). The second term is the $q_l (i)$ where $i$ is a water molecule, and only water molecules are considered as neighbors when performing the calculation. The last one is the parameter for an alcohol molecule, only considering alcohols molecules as neighbors. 

Finally, we have a vector $\mathbf{q}(i)$ for each particle $i$, composed by all the different BOOPs and cubic BOOPs (and their averages), taking into consideration the variations applied for the binary-mixture case. For any molecule we have

\begin{eqnarray}
 \mathbf{q} (i) = \Big( \big\{\textit{q}_{l}^{W,A} (i) \big\}, \big\{ \bar{q}_l^{W,A} (i) \big\}, \big\{ \bar{\bar{q}}_l^{W,A} (i) \big\},
 \big\{ q_l^{W,W} (i) \big\},\nonumber \\
 \big\{ \bar{q}_l^{W,W} (i) \big\}, 
 \big\{ \bar{\bar{q}}_l^{W,W} (i) \big\},  \big\{ q_l^{A,A} (i) \big\},  \big\{ \bar{q}_l^{A,A} (i) \big\}, \big\{ \bar{\bar{q}}_l^{A,A} (i) \big\},\nonumber \\  \big\{\textit{w}_{l}^{W,A} (i) \big\}, 
\big\{ \bar{w}_{l'}^{W,A} (i) \big\},\big\{ \bar{\bar{w}}_{l'}^{W,A} (i) \big\}, \big\{ w_{l'}^{W,W} (i) \big\}, \big\{ \bar{w}_{l'}^{W,W} (i) \big\}, \nonumber \\
 \big\{ \bar{\bar{w}}_{l'}^{W,W} (i)  \big\},
 \big\{ w_{l'}^{A,A} (i)  \big\},  \big\{ \bar{w}_{l'}^{A,A} (i) \big\}, \big\{ \bar{\bar{w}}_{l'}^{A,A} (i) \big\} \Big)
 \label{q_vector}
\end{eqnarray}

with $l \in [3,12]$ and $l'$ assuming only the even values of $l$. For a water molecule, all the parameters with \textit{A,A} in the exponent are equal to zero. If we consider an alcohol molecule, we change \textit{W,A} for \textit{A-W} in (\ref{q_vector}) and set all parameters with \textit{W,W} in the exponent to zero.

Thus, $\mathbf{q} (i)$ is a 135-dimensional vector that uniquely identify the local structure of the systems' particles up to the third-shell neighbors, distinguishing water molecules from alcohol molecules.

The model chosen to autonomously identify the local environments is a Feed-Forward Neural Network, with the vector in (\ref{q_vector}) as the Input Layer (IL), three Hidden Layers (HL) with 180, 90 and 30 neurons, respectively, and a 5-dimensional Output Layer (OL), where each one of the output neurons represents the probability of a particle being in one of the five possible phases the systems analyzed can assume. The whole NN approach was performed using the keras \cite{chollet2015keras} package, with TensorFlow \cite{tensorflow2015-whitepaper} backend. Glorot initialization \cite{pmlr-v9-glorot10a} was applied for all layers; Rectifier Linear Unit was used as activation function for the IL and the HLs, whereas Softmax was used as activation function for the OL; sparse categorical crossentropy was used as loss function with adam \cite{kingma2015adam} as optimizer. The training was performed for 40 epochs, using a batch size equals to 32. The choice to use a NN as the ML model used to perform the classification is based on the constant success obtained when using this specific model, such as in \cite{boattini2018, martelli2020, Gartner26040}. There's nothing special about the model specifications, different network's architecture were tested, as well as hyperparameters, and the setting described presented the best validation accuracy overall.

To efficiently teach the network to distinguish the different phases, we used as input a $\mathbf{q}(i)$ vector for each CSW particle of a system, from a total of thirty configurations, two for each phase and for each mixture with alcohol concentration equals to $10\%$. Since the number of particles per system is equal to 1000, the total number of data points for the train set is 30 thousand, each one labeled with one of the five possible phases. Since the snapshots with the particle's positions are taken from systems which are in equilibrium, we assume it is uniform, with all the N particles in the same phase. So, the phase used for label train data is already known, as it was found from the thermodynamic analysis performed in our previous work.\cite{marques2021} For example, if we use the N particles from the ethanol-water mixture with concentration $\chi = 0.10$, for temperature $T = 0.20$ and $P = 0.01$, as part of the train set, we use our previous analysis to label the N particles with `phase I'. Increasing  the number of snapshots used for training showed no relation with better results overall. The impact that the configuration uniformity had on the accuracy was much influential than the number of configurations composing the train set.

Afterwards, we use the trained NN to predict the phase of the particles from all the configurations analyzed in \cite{marques2021} i.e. the three different mixtures with three different concentrations, besides pure water. The results regarding the predicted phase for each particle give two different information. The first one is the system's phase, which is simply found by analyzing the dominant phase in a single system. The second one is the type of phase transition occurring near the transition curves, which can be extracted from a population analysis (the number of particle in each phase for a single system) . With the thermodynamic analysis, transitions points occur where response functions present a maximum or a discontinuity. First and second order phase transitions can be distinguished analysing the response functions, which result in the transitions points represented in Figure \ref{fig:dfs-etha-and-water} as grey points. Moreover, the population behavior can give more insights about the system phase behavior in the supercooled regime.

\section{Results and discussion}

The neural network achieved accuracy up to $99\%$ and $99.3\%$, for training (see Supplementary Material - SM) and validation, respectively. Similar results can also be achieved with one less hidden layer, but increasing the epochs. Overall, it was the faster method to maintain a higher number of hidden layers, but train with less epochs. 

\begin{figure}[h!]
\begin{tabular}{c}
  \includegraphics[width=0.45\textwidth]{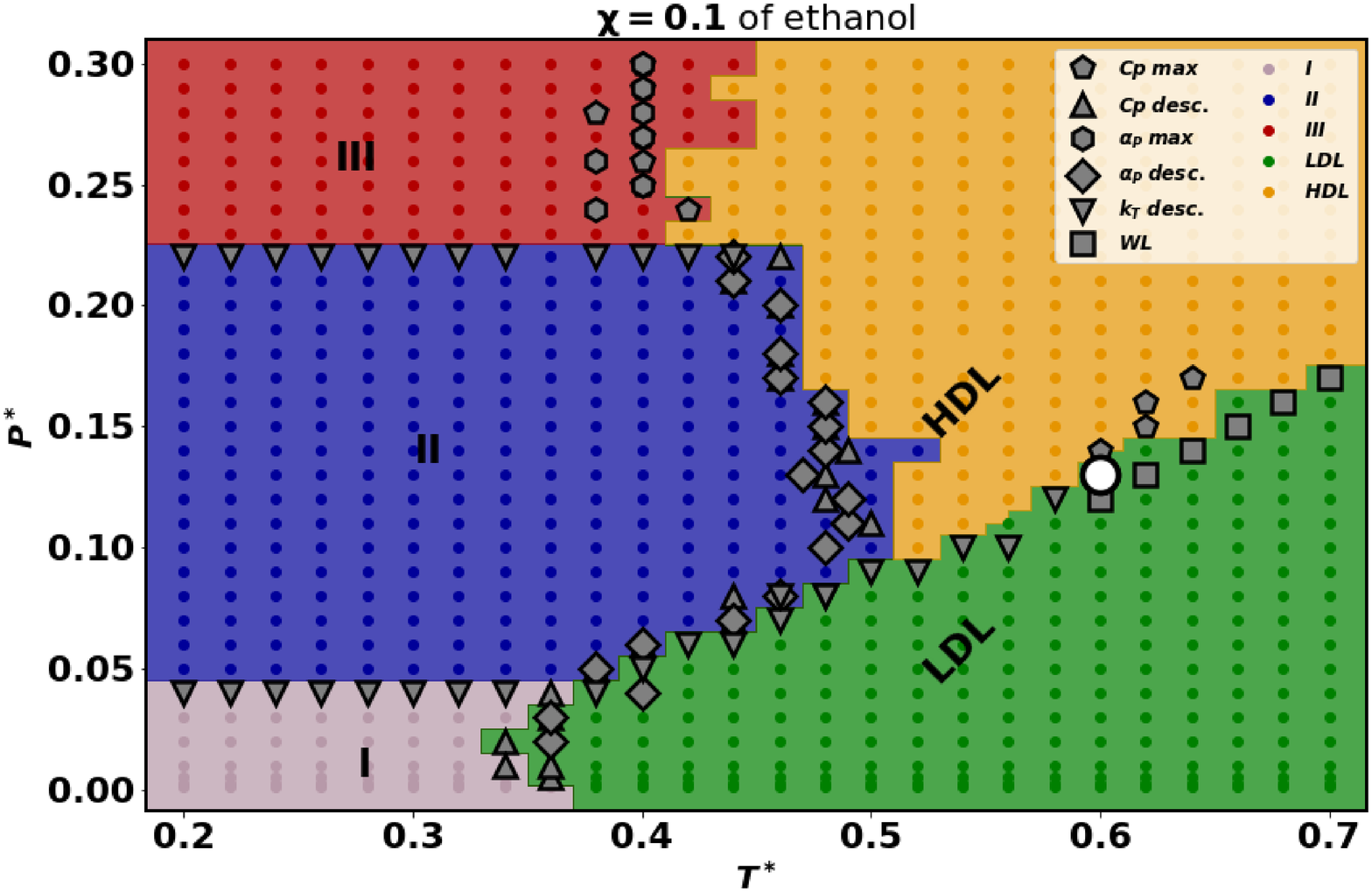} \\ (a) \\  \includegraphics[width=0.45\textwidth]{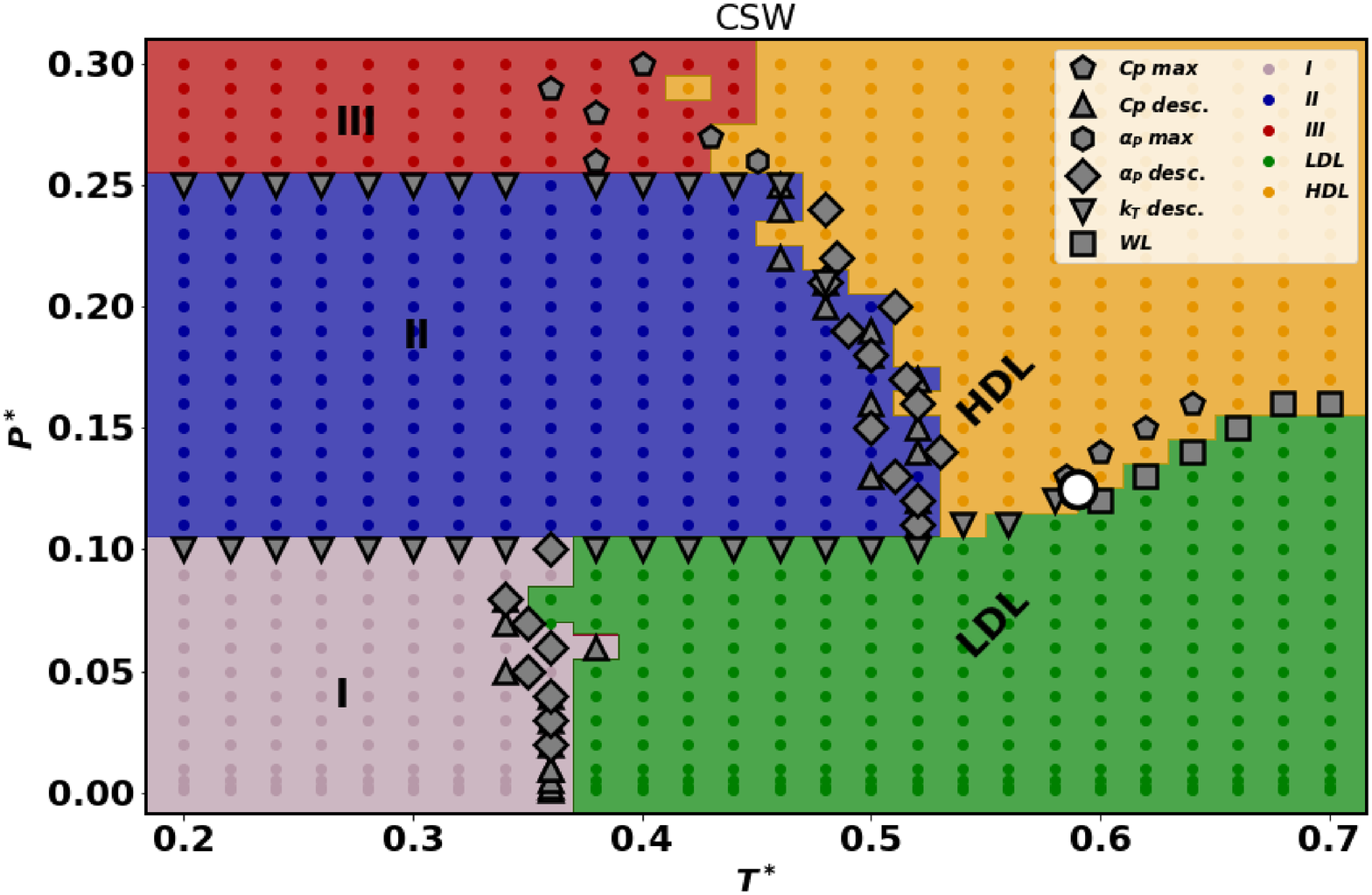} \\ (b)  \\[6pt]
\end{tabular}
\caption{Phase diagrams of (a) water-ethanol mixture with concentration $\chi = 0.1$ and (b) pure water, predicted by the neural network. The grey region points were classified as solid phase I, blue region dots as solid phase II, red region dots as amorphous phase III, green region dots as LDL and the yellow region dots as in the HDL phase. The grey markers with distinct shapes stands for discontinuities and maxima in the response functions isothermal compressibility $\kappa_T$, isobaric expansion coefficient $\alpha_P$ and specific heat at constant pressure $C_P$ taken from Ref.~\cite{marques2021}. The white circle is the LLCP obtained from the Ref.~\cite{marques2021}.}
\label{fig:dfs-etha-and-water}
\end{figure}

    In Figure \ref{fig:dfs-etha-and-water}(a) is presented the phase diagram obtained with the neural network for the water-ethanol mixture with concentration of ethanol equals to $10\%$. This is one of the systems from which two snapshots per phase were used to train the NN. Here, each colored circle corresponds to one configuration with defined pressure, temperature and phase, being each phase represented by a unique color. The grey markers correspond to the transition points found from the thermodynamic analysis, forming the ``ground-truth'' transition lines that delimit each phase and the white circle is the LLCP. We notice from the figure that there is a great overall correspondence between the phases predicted by the network and by the classical approach. That behavior is also noticed for pure CSW, Figure \ref{fig:dfs-etha-and-water}(b), and all the combinations of mixture and concentration, shown in the SM. The overall accuracy, defined as the number of configurations with defined temperature and pressure that the network correctly predicted the phase divided by the total number of configurations, are shown in Table 1. Here we note that the overall accuracy found when predicting the phase diagrams is significantly lower than the train/validation accuracy. That is explained by the configurations near the transitions lines which phases were incorrectly classified by the network. Also, as it can be seen in the Supplementary Material, the accuracy is already high in the first epochs of the training process, and then increases gradually, indicating that the network learns the main model's features quickly. Also, is important to address that the randomness in choosing the systems for the train dataset plays a major role in the accuracy. For instance, if a random point near the phases boundaries is chosen, the agreement is worse since in these regions we can have a mixture of two phases, what leads to a bad train set. 
    \begin{table}[h!]
\caption{Accuracy for pure water and all combinations of mixtures.}
\begin{center}
    
\begin{tabular}{lcc}
&$\chi$&Accuracy\\
\hline
& &\\
CSW&-&0.94\\
& &\\
\hline
&0.01&0.93\\
Methanol&0.05&0.92\\
&0.10&0.94\\
\hline
& 0.01 & 0.89 \\
Ethanol& 0.05 & 0.92 \\
& 0.10 & 0.93 \\
\hline
   & 0.01 & 0.90 \\
 Propanol  & 0.05 & 0.94 \\
   & 0.10 & 0.92 \\
\end{tabular}
\end{center}
\end{table}

\begin{figure}[b]
\begin{center}

\begin{tabular}{c}
 \includegraphics[width=0.45\textwidth]{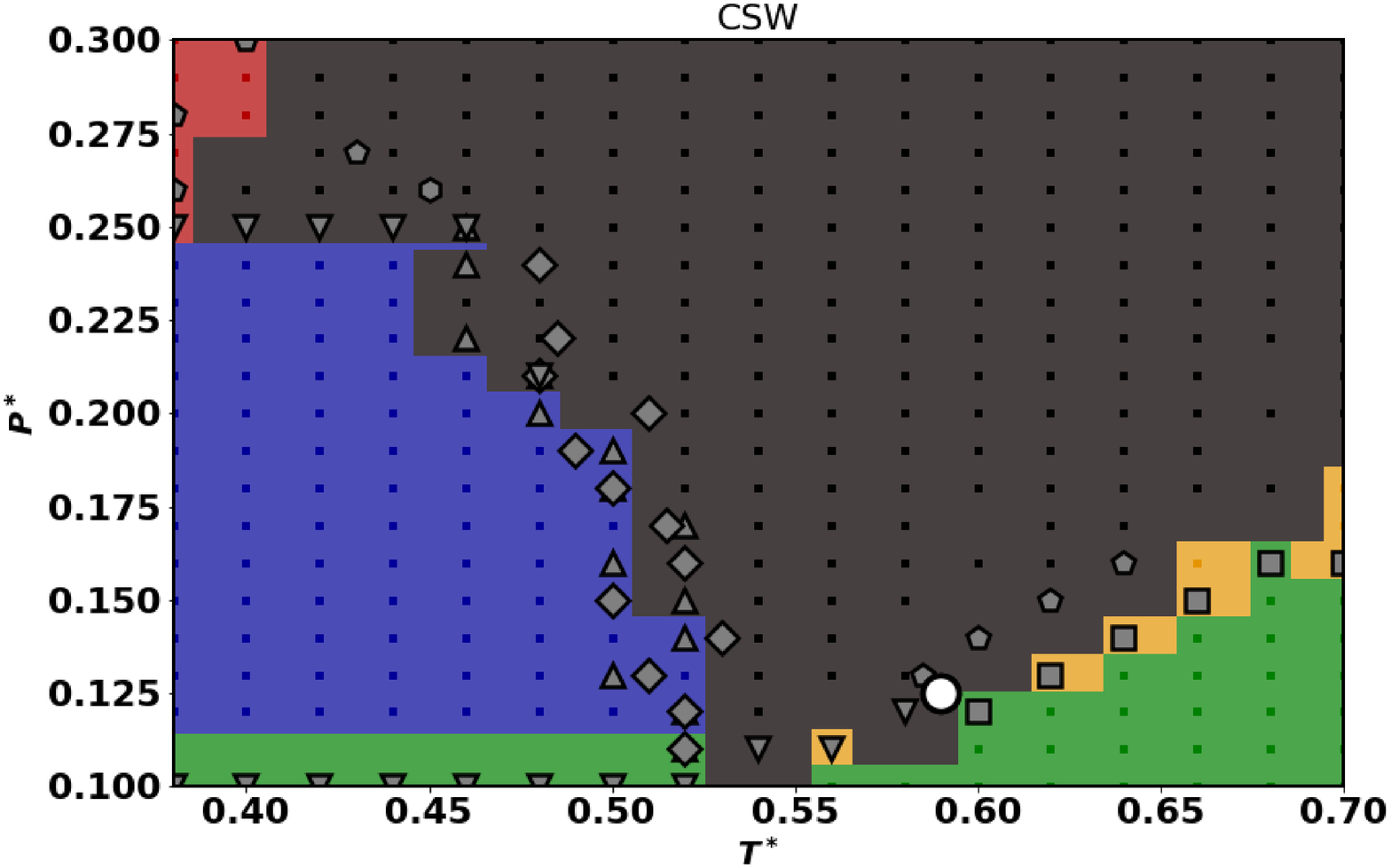}\\
 (a)\\
 \includegraphics[width=0.45\textwidth]{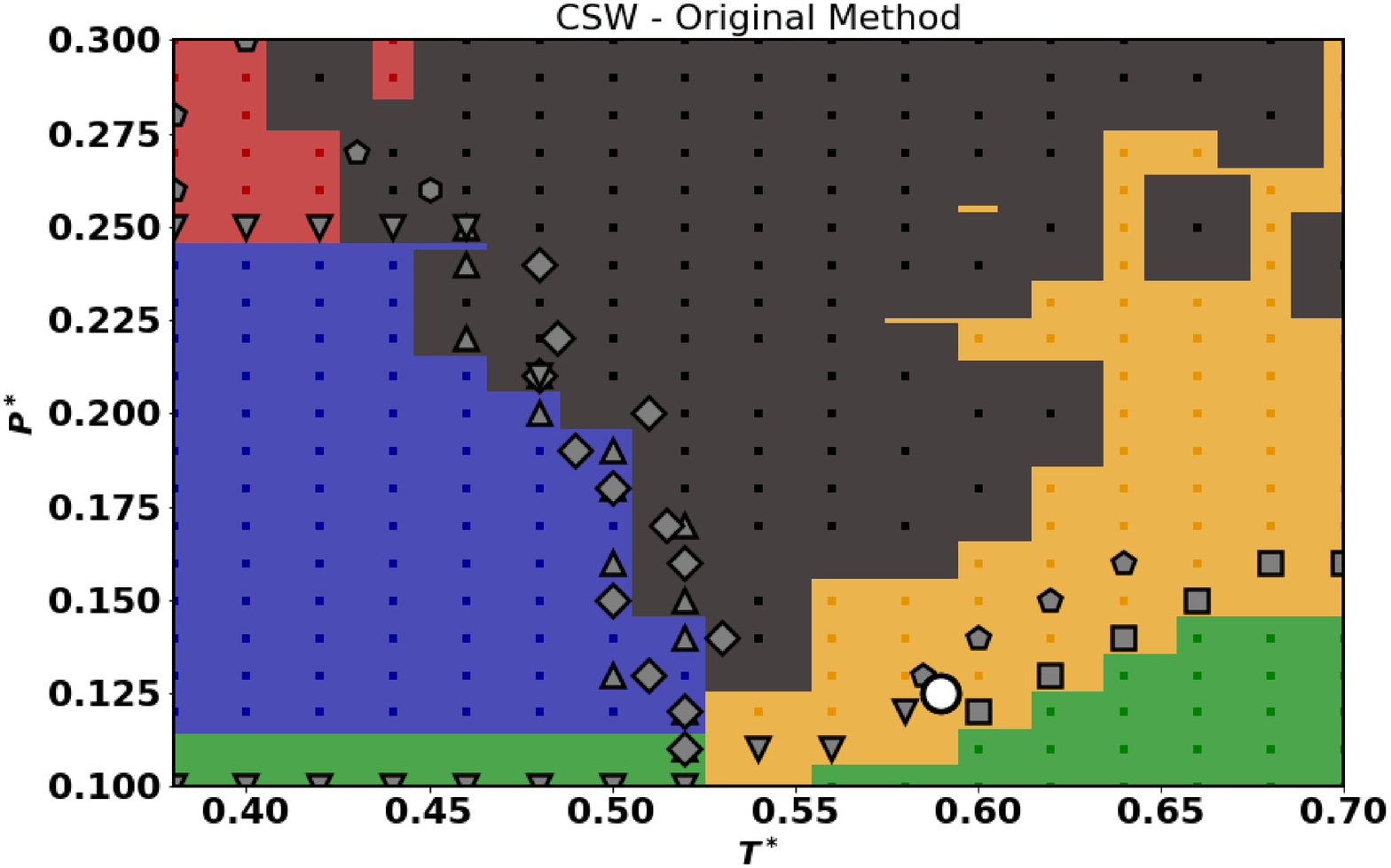}     \\
 (b)  \\
\end{tabular}
\end{center}

\caption{
Phase diagrams predicted using BOOPs and their average 
(a) and including the average-average parameters (b) for the pure CSW fluid. Blue dots were classified as phase II, red as phase III, green as LDL, yellow as HDL and the black dots indicates the existence of amorphous-like particles in the HDL phase.}
\label{fig:metaswater}
\end{figure}

 \begin{figure}[t]
 \begin{center}
\begin{tabular}{c}
\includegraphics[width=0.45\textwidth]{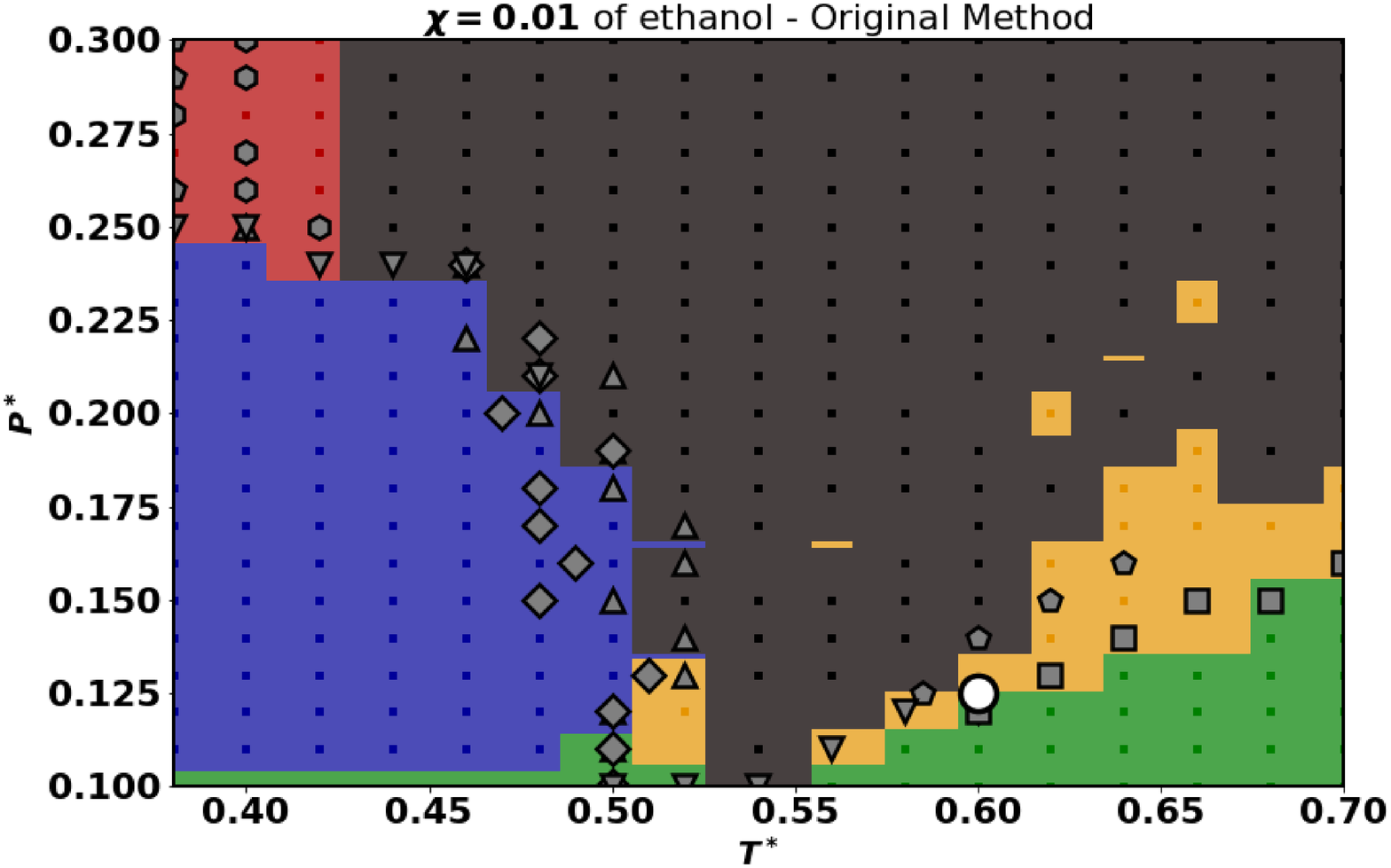} \\
(a) \\
\includegraphics[width=0.45\textwidth]{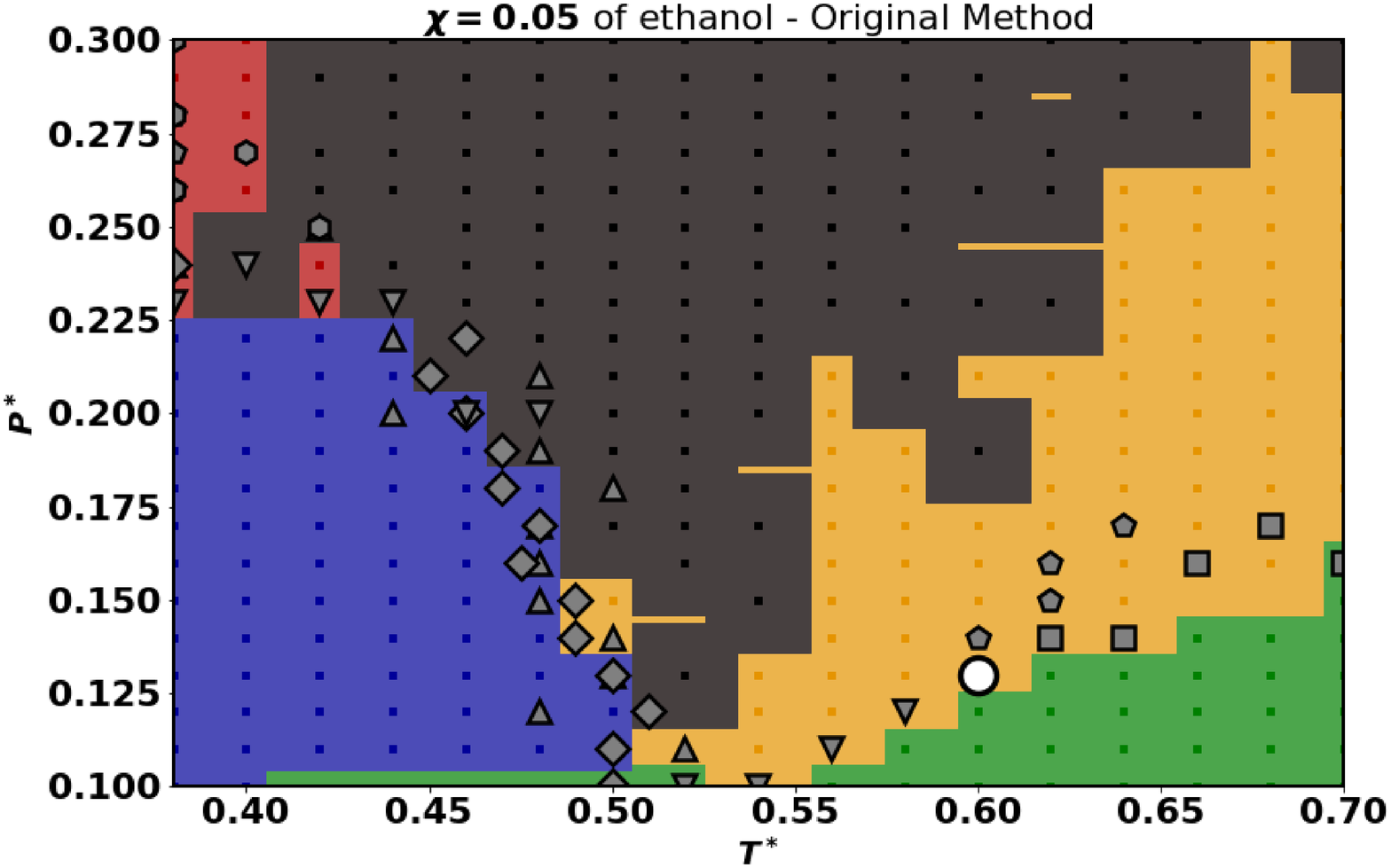} \\  
(b) \\
\includegraphics[width=0.45\textwidth]{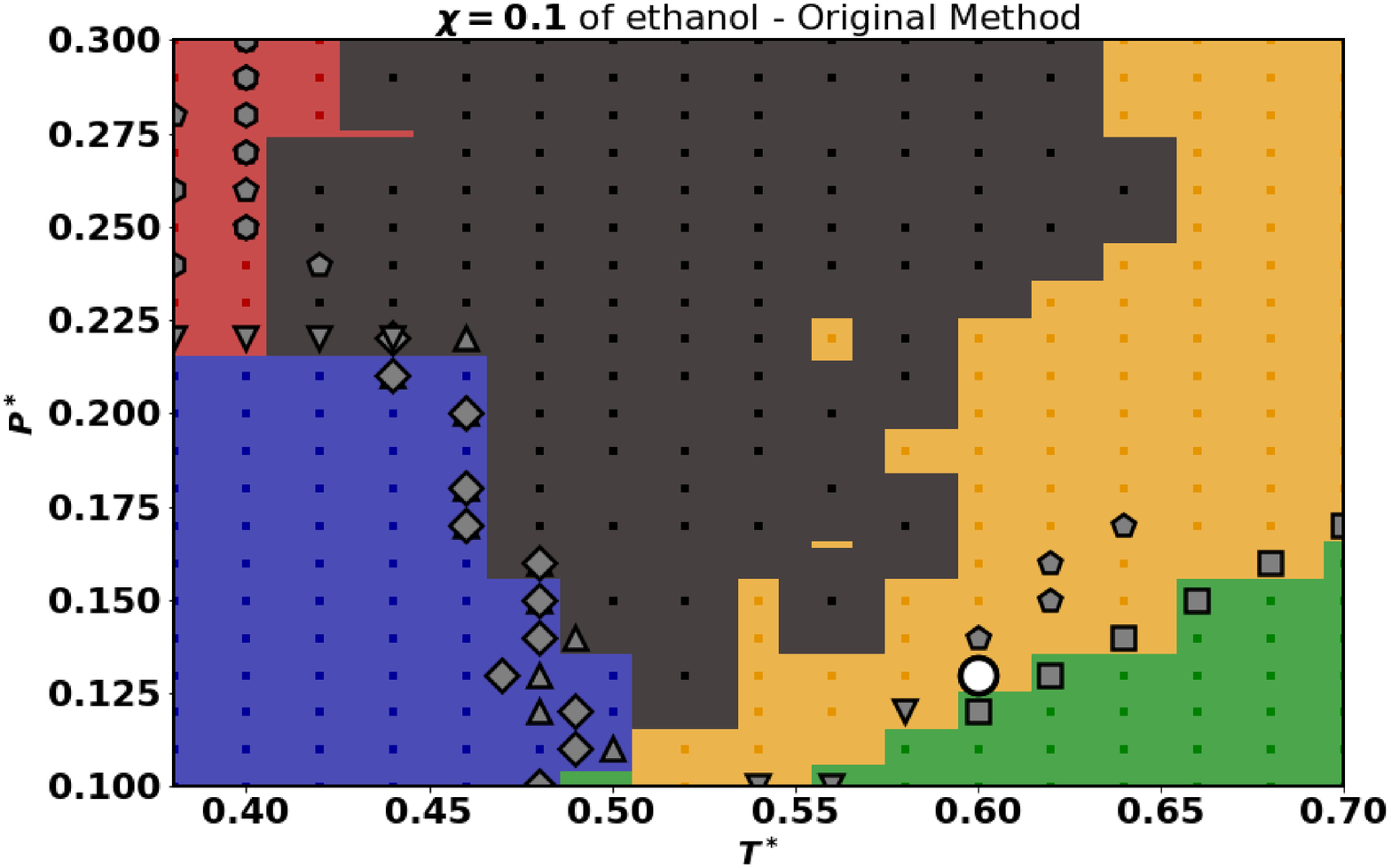}   \\
(c)\\
\end{tabular}
\end{center}
\caption{ Phase diagrams predicted by the NN using average-average parameters, for $0.38 \leq \mathbf{T^*} \leq 0.70$ and $0.10 \leq \mathbf{P^*} \leq 0.30$ of water-ethanol mixture with (a) $\chi = 0.01$, (b) $\chi = 0.05$ and (c) $\chi = 0.1$.  Blue dot regions were classified as phase II, red regions as phase III, green as LDL, yellow as HDL and the black region is the metastable phase (phase classified as LDL with particles classified as phase III). }
\label{fig:metas}
\end{figure}

    The compatibility between the methods is particularly interesting in the supercritical region, for temperatures higher than the critical temperature. In this region, we cannot precisely define two different liquid phases, but one liquid phase with two different characteristics, one closely related to the HDL phase and the other to the LDL phase. The separation within supercritical liquid phase is given by the Widom Line (WL), an extension of the LLPT curve into the one-phase region and the locus of maximum fluctuations of the order parameter \cite{holten2012, bianco2019}, represented by the grey squares in the diagram. The NN successfully separates the HDL-like supercritical liquid from the LDL-like one, and the HDL and LDL liquids in the subcritical region. To check if the system is crossing the phase coexistence line or the WL, we can analyze the isothermal populations. This population analysis is also useful to analyze the amorphous-HDL boundary region, where the larger discrepancy between the neural network approach and the thermodynamic analysis was observed. At this point, is important to recall that the amorphous phase is not an equilibrium one, and the amorphous-HDL boundary may change if the system is going through a cooling or a heating process -- our results were evaluated by a cooling process. The amorphous phase has a smaller diffusion constant -- $D\approx 0$ -- compared to the HDL phase, what indicates absence of movement  - for this reason, we are calling it a ``solid". Also, the maxima in $\alpha_P$ and $C_P$, shown in the phase diagrams (Figure \ref{fig:dfs-etha-and-water}) and the smooth change in the structural parameters, as the pair excess entropy $s_2$, suggest a boundary between the amorphous and HDL phases \cite{marques2021}. Once the HDL and the amorphous phase have similar short-range ordering \cite{marques2021}, we expand the method by Martelli and co-authors \cite{martelli2020} to include structural information about the third-shell neighbors, still considering different parameters for different molecule, as done by Boattini \textit{et al} \cite{boattini2018}. It is important for our case once the waterlike characteristics of core-softened fluids can be related to competitions in the long range coordination shells - not only in the first or second one.\cite{Krek08,franzeseJCP2010, marques2021} To this end, we include the average-average therms, given by (\ref{BOOP2c}) and (\ref{BOOP3c}). This approach leads to a slightly better agreement between the NN method and the analysis based in the response functions. The overall accuracy (the number of configurations for which the phase was correctly predicted) do not improve much, however, the results are significantly better when we look at the population of particles in a particular phase for each point in the phase diagram. 
    
\begin{figure*}[t]
\begin{tabular}{cc}

  \includegraphics[width=0.49\textwidth]{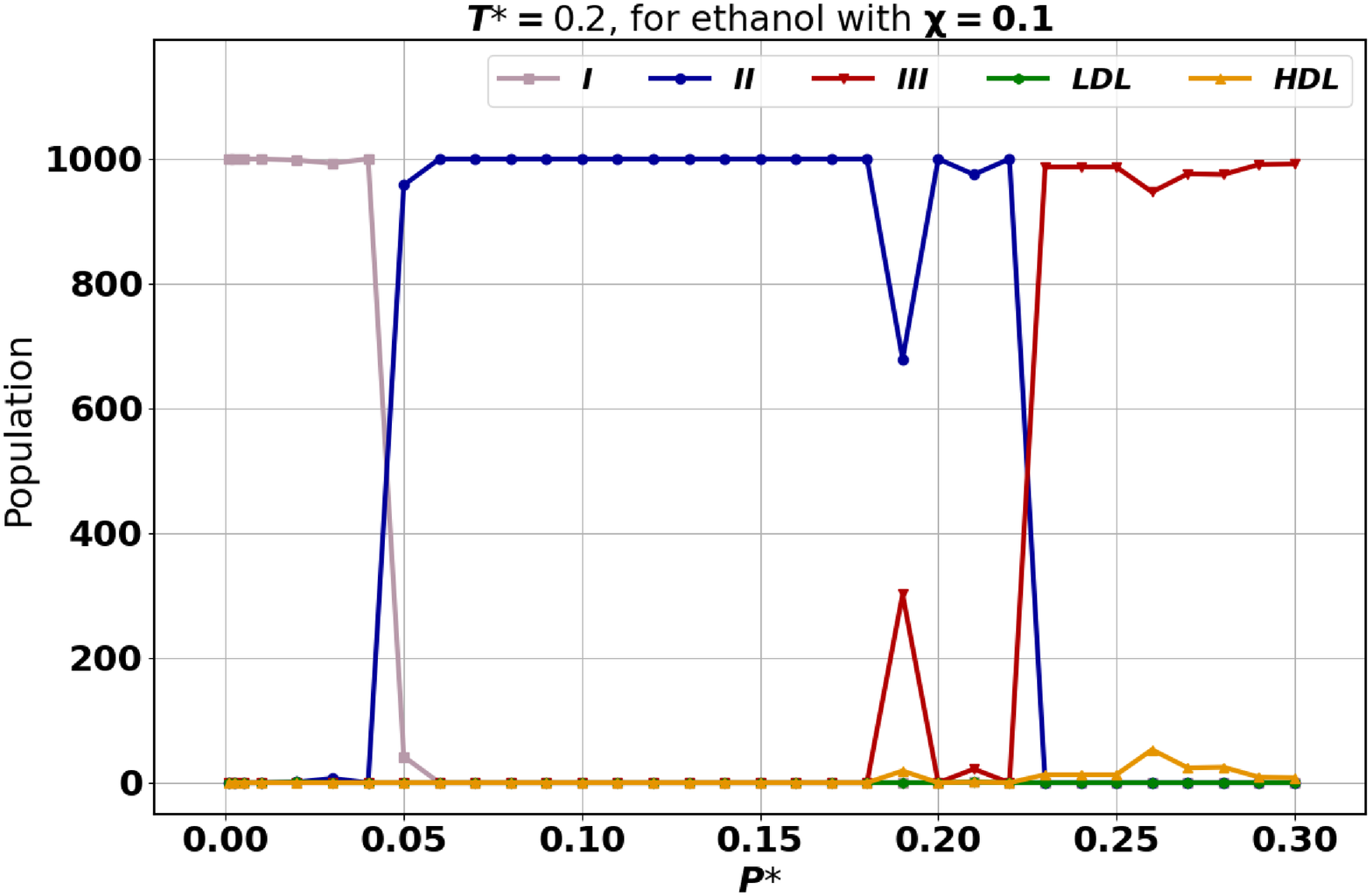} &   \includegraphics[width=0.49\textwidth]{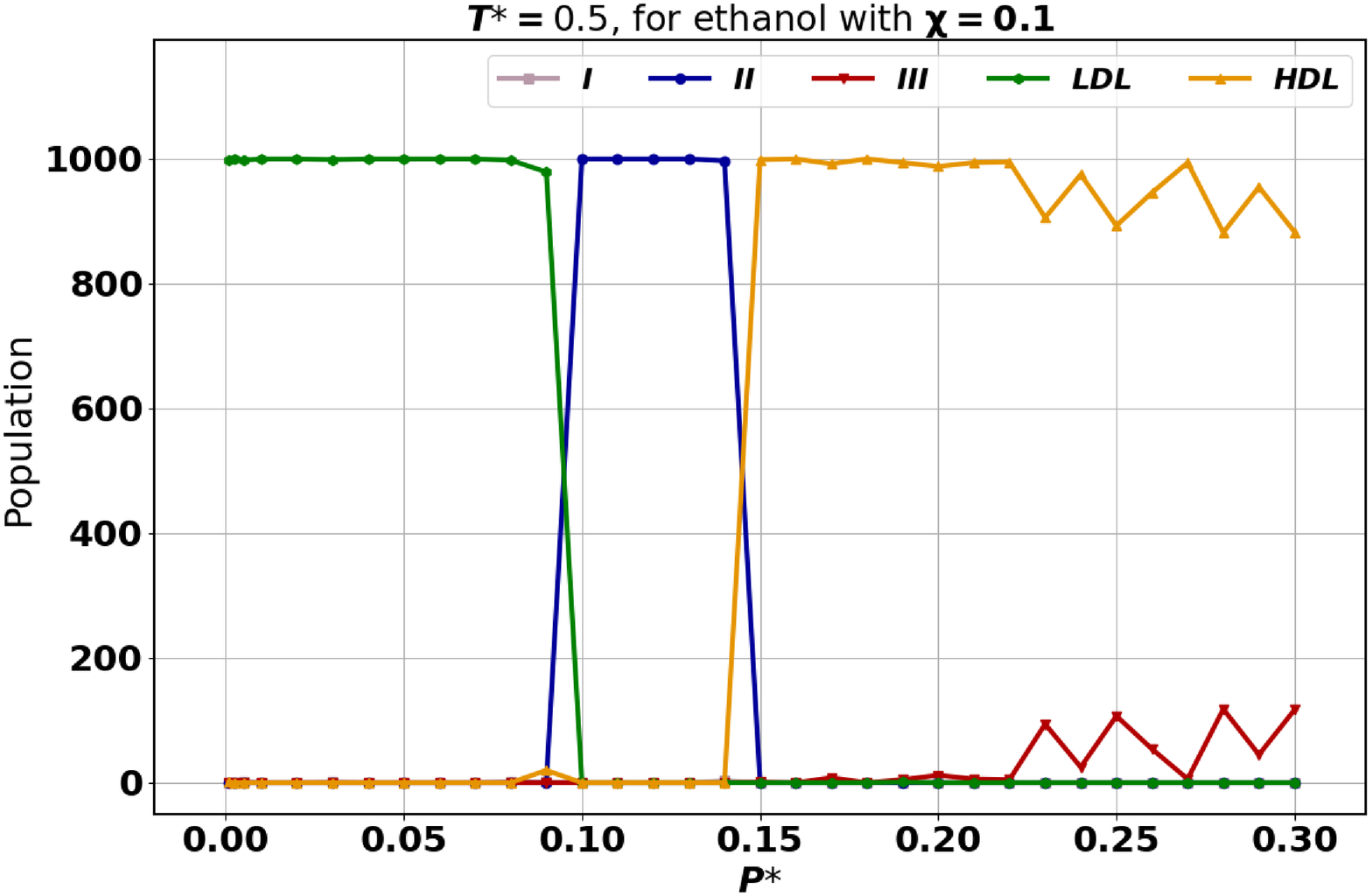} \\
(a)  & (b)  \\[6pt]

 \includegraphics[width=0.49\textwidth]{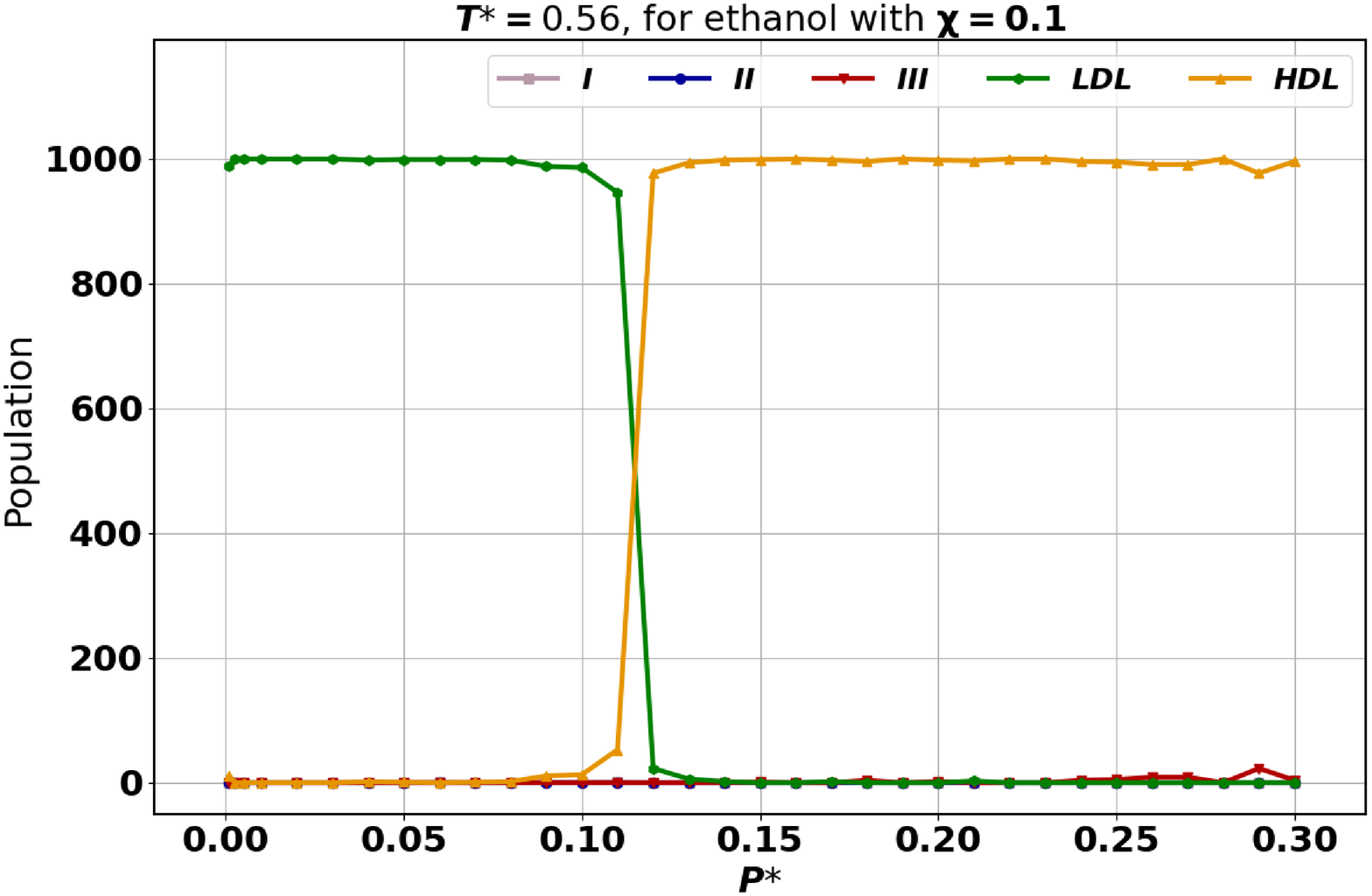} &   \includegraphics[width=0.49\textwidth]{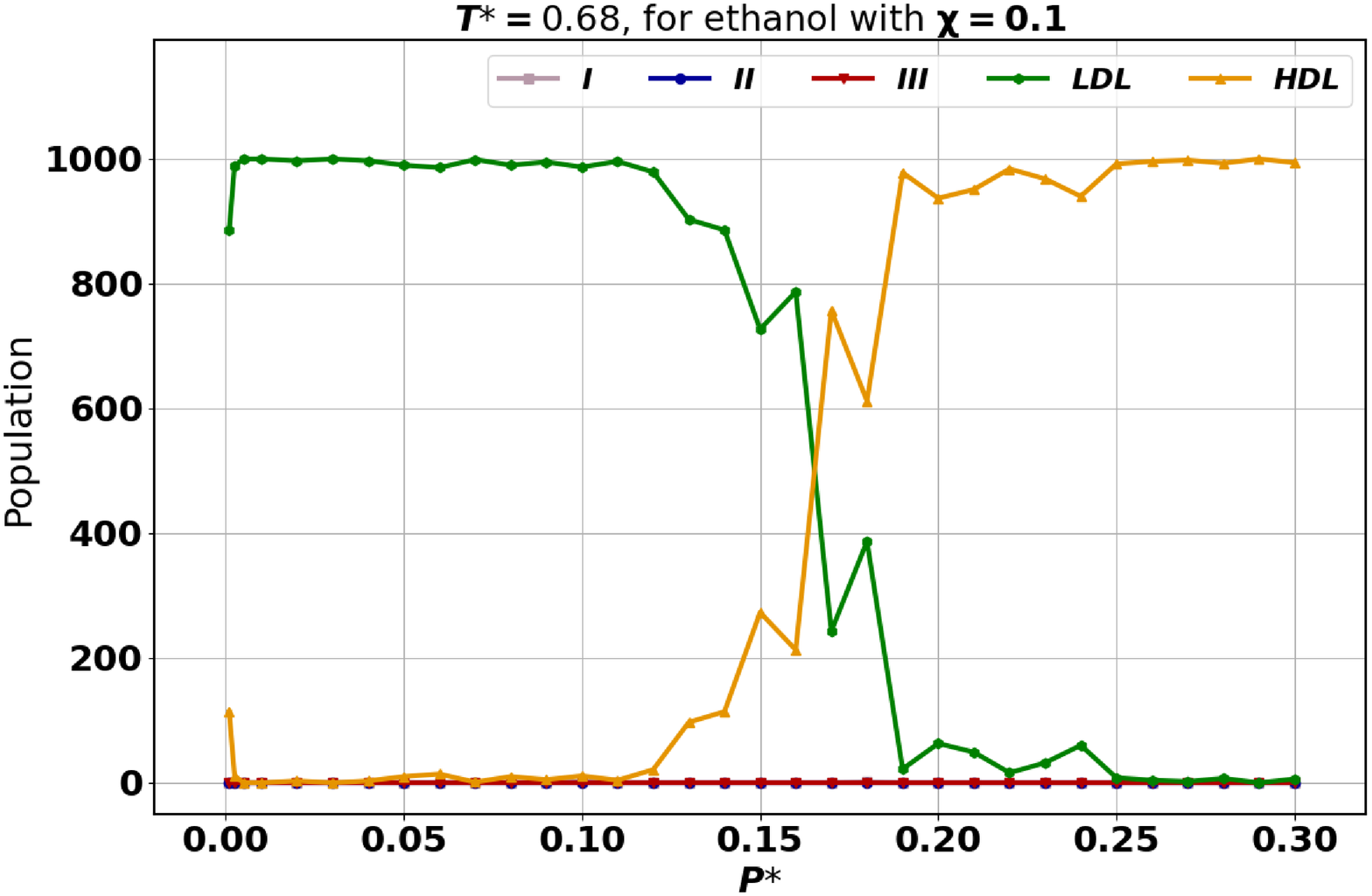} \\
(c)  & (d)  \\[6pt]
\end{tabular}
\caption{Population as a function of pressure for the water-ethanol mixture with concentration equals to 0.1 and for fixed temperature equals to (a) 0.2, (b) 0.5, (c) 0.56 and (d) 0.68.}
\label{fig:population-p-ethanol-0.1}
\end{figure*}

 \begin{figure*}[t]
\begin{tabular}{cc}

  \includegraphics[width=0.49\textwidth]{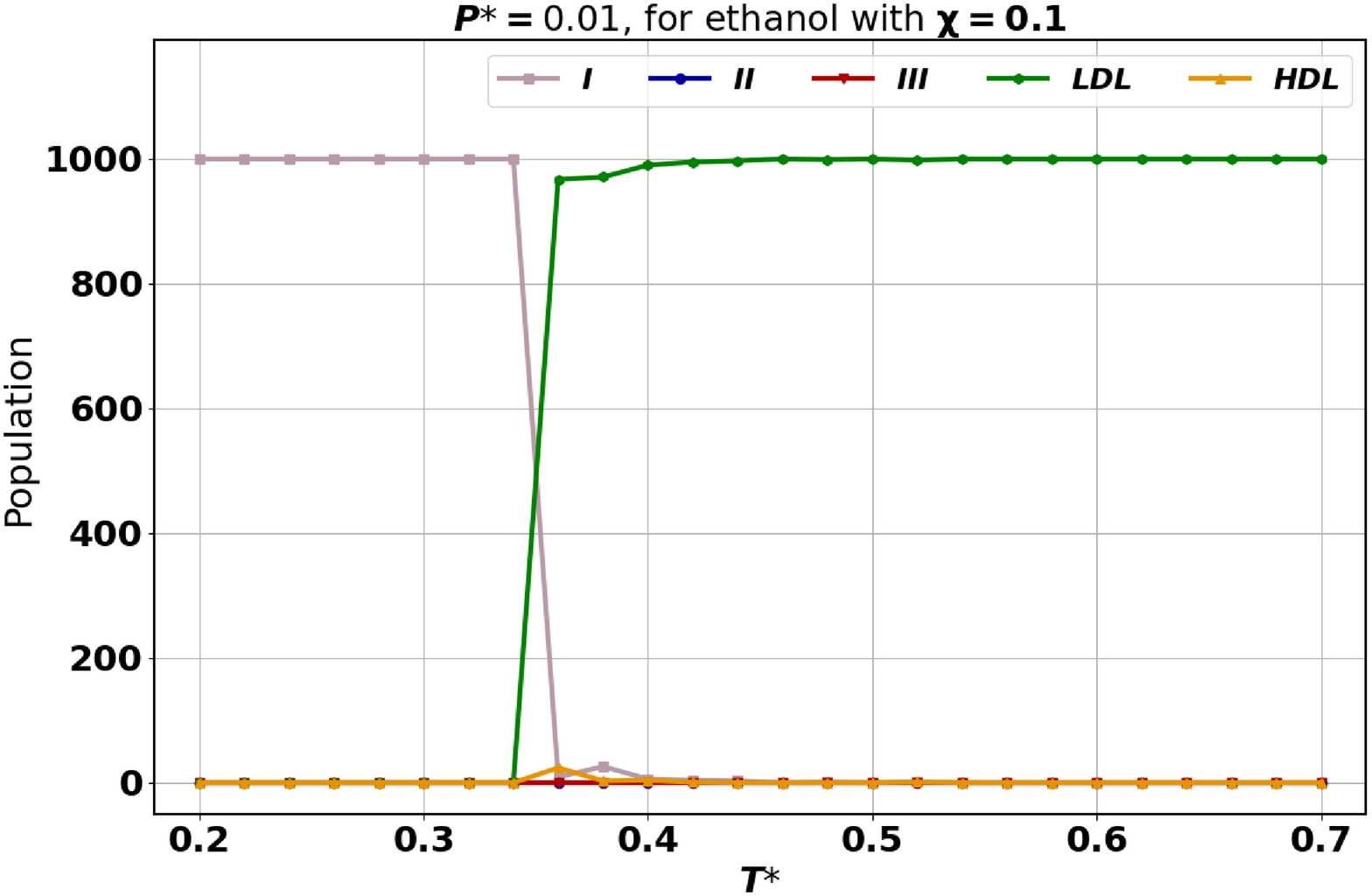} &   \includegraphics[width=0.49\textwidth]{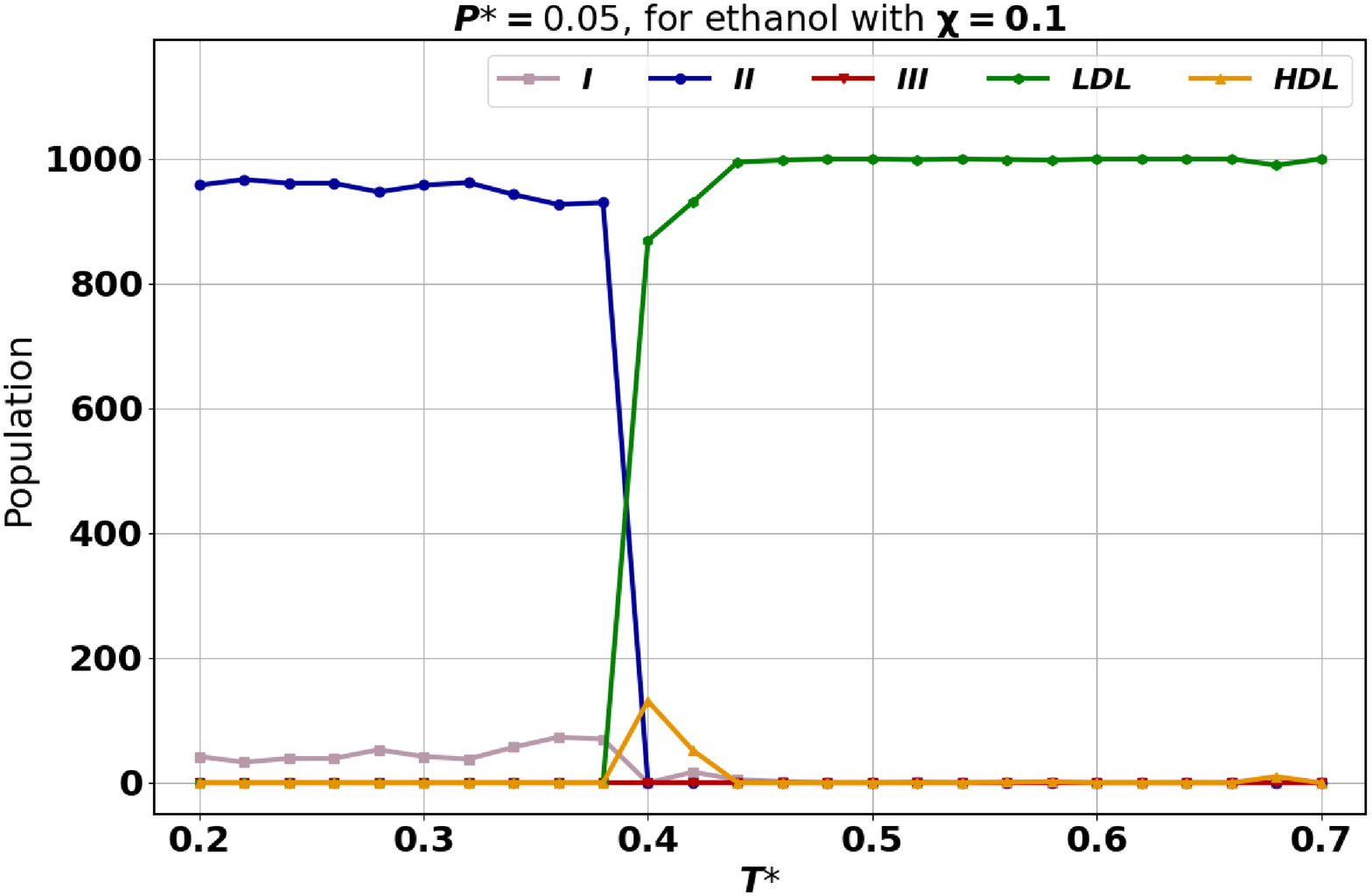} \\
(a)  & (b)  \\[6pt]

 \includegraphics[width=0.49\textwidth]{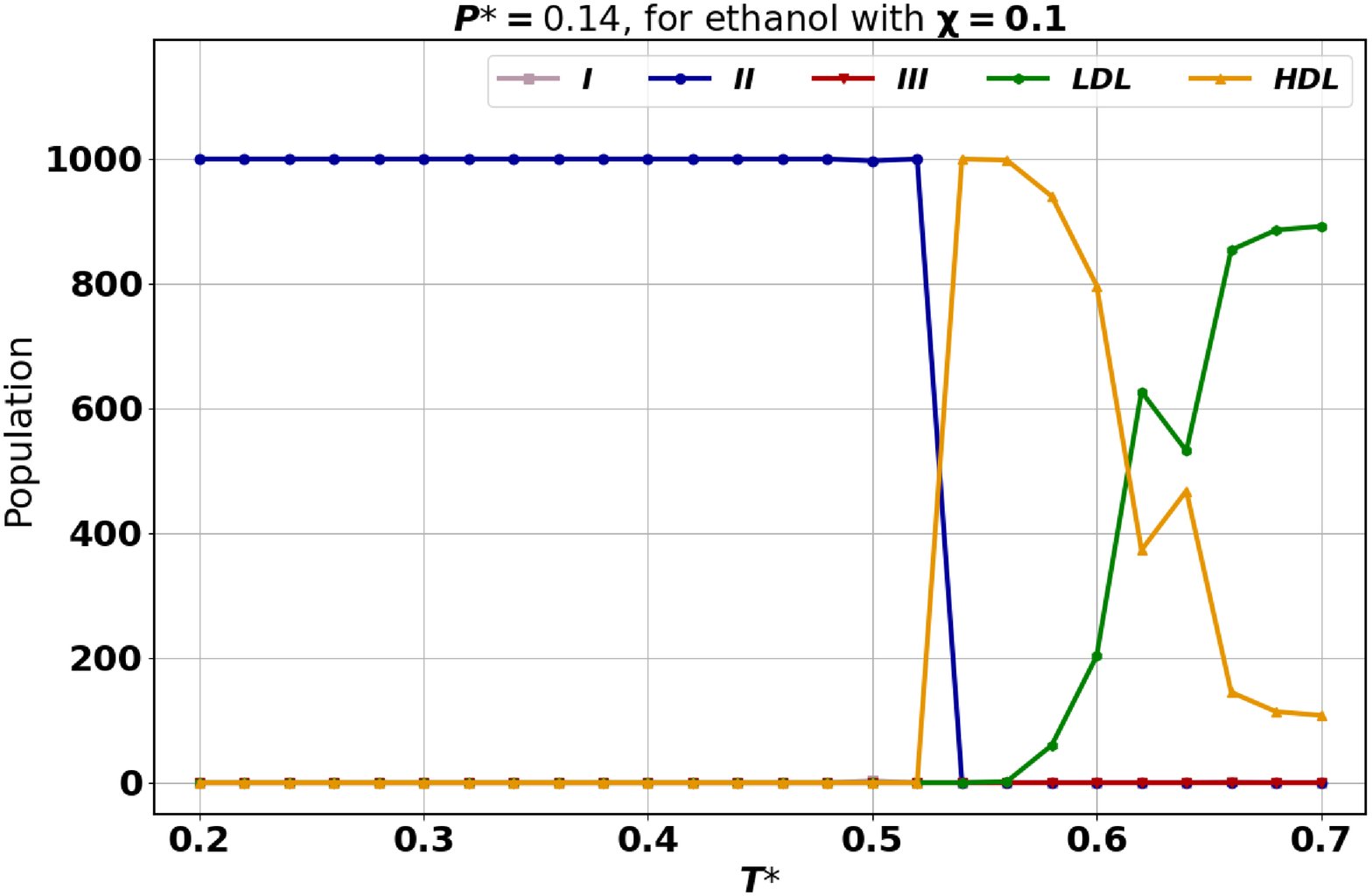} &   \includegraphics[width=0.49\textwidth]{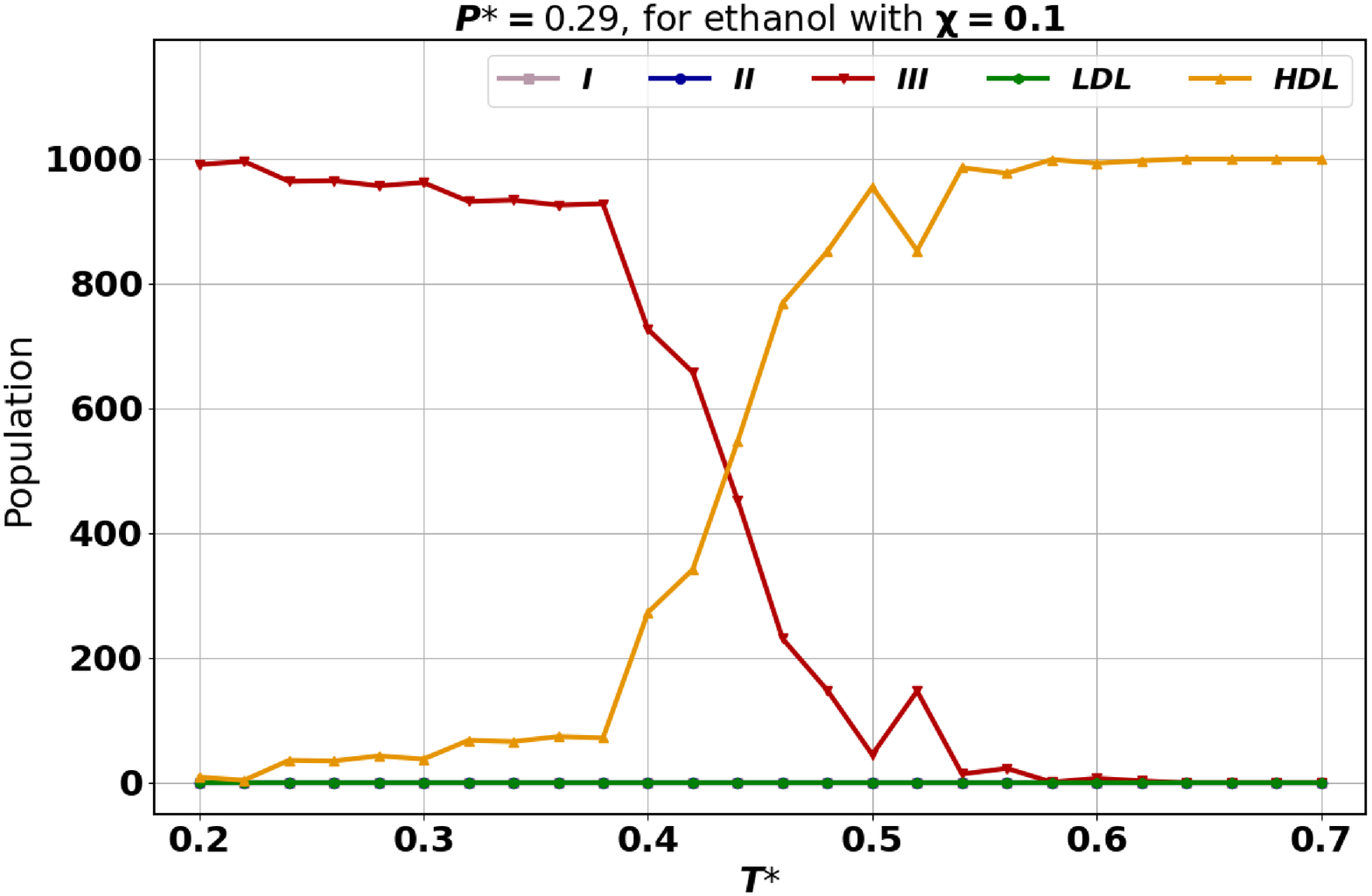} \\
(c)  & (d)  \\[6pt]
\end{tabular}
\caption{Population as a function of temperature for the water-ethanol mixture with concentration equals to 0.1 and for fixed pressures equals to (a) 0.01, (b) 0.05, (c) 0.14 and (d) 0.29.}
\label{fig:population-t-ethanol-0.1}
\end{figure*}

Each point in the phase diagram was defined using the population analysis: if 50\% or more of the particles are in one of the five possible phases, per say HDL phase, the point is classified as HDL. Nevertheless, some particle in the HDL phase can have populations classified as a distinct phase. As an example, we evaluate the population of amorphous particles in points that were classified in the stable HDL phase, since it can indicate how the metastable region spreads in the phase diagram. In Figure \ref{fig:metaswater} (a) and (b) we compare the case of pure CSW water using 
previously implemented methods \cite{boattini2018, martelli2020}, adapted to our system,
and our version, that includes longer correlations. 
The points in the metastable phase -- which were classified as HDL but have at least one particle classified as amourphous, 
are painted black, in contrast with the colors of the Figure \ref{fig:dfs-etha-and-water}(b). While in the first phase diagram amorphous particles spread along practically every point in the HDL phase, in the second one the amorphous population occupy a smaller region. This indicates that including longer-range information will lead to a better classification of these glassy phases.

We can also apply this analysis -- with the average-average parameters -- to see how the amorphous-like population changes as the concentration of alcohol in the solution increases. As we can see in Figure \ref{fig:metas}(a) for CSW-ethanol mixtures, and in the SM for the other alcohols, the region with amorphous population in the HDL phase increases for the lower concentration in comparison to the pure CSW fluid, Figure \ref{fig:metaswater}(b), and then shrinks with the increase in concentration, as shown in Figures \ref{fig:metas}(b) and (c). This agrees with our previous results \cite{marques2021}, where we found that the low concentration of alcohol affects only the long range coordination shells. Then, as $\chi$ increases, it favors the short range organization, that becomes predominant and the long range effects are less relevant.

   Using the NN classification, we can define how much particles are in one of the five phases. The number of particles defines the population of each phase. For instance, we can walk along an isotherm and see how the population changes. Taking the low temperature isotherm $T^* = 0.20$ for the ethanol mixture at $\chi = 0.1$, shown in Figure \ref{fig:population-p-ethanol-0.1} (a), we can see that at the I-II transition practically all particles change from the BCC to the HCP structure. However, at the II-III transition, we can see fluctuations in the II and III populations from $P^* = 0.19$ to $P^* = 0.22$. Once this corresponds to a solid-amorphous transition, this is expected due the metastability of the phase III. After that, we can see the presence of HDL-like particles in the region III. Heating to $T^* = 0.50$, Figure \ref{fig:population-p-ethanol-0.1} (b), an isotherm that crosses liquid and solid phases, we can see the LDL-solid II transition at $P^* = 0.10$ by the abrupt change in the populations of each one of the phases, same for the solid II-HDL transition at $P^* = 0.15$. Once again, the effect of the amorphous phase metastability has been noticed: at higher pressures the population of amorphous particles starts to increase. At the subcritical isotherm $T^* = 0.56$, shown in Figure \ref{fig:population-p-ethanol-0.1} (c), the LDL-HDL transition is clear at $P^* = 0.12$ -- distinct from the supercritical isotherm $T^* = 0.68$, shown in Figure \ref{fig:population-p-ethanol-0.1} (d), where the transition from LDL-like to HDL-like behavior is continuous. Also, this temperature is high enough -- and far enough from the metastable region -- to ensure that there is no more amorphous-like particles in the system.

A similar analysis can be made along isobars. Here, we show the isobars $P^* = 0.01$, $P^* = 0.05$, $P^* = 0.14$ and $P^* = 0.29$ in Figure \ref{fig:population-t-ethanol-0.1}(a) to (d), respectively. The first two show the abrupt change in solid and liquid populations for the solid I - LDL and solid II - LDL transitions. In the Figure \ref{fig:population-t-ethanol-0.1} (c) we can see the solid phase II to HDL transition at $T^* = 0.51$ followed by a change in the HDL-like and LDL-like populations as the isobar crosses the WL. Finnaly, we observe that the amorphous - HDL phase transition is smooth, with the particles structure gradually changing from one type to another until the high temperature limit, where no more fluctuations from the metastable phases are observed.

\section{Conclusion}

In this paper, we have used a machine learning approach to classify phases of core softened CSW-alcohol mixtures, for different alcohols and concentrations, as well as pure CSW, in the supercooled regime. The neural network model inspired by 
 \cite{boattini2018} and \cite{martelli2020} uses an extensive set of unique bond-orientational order parameters for water-water, water-alcohol and alcohol-alcohol bonds, as input, and was extended to include longer-range coordination shells in comparison to the original method.

For pure CSW fluid and for all the possible combinations of mixtures and concentrations, the phase classification agrees with the thermodynamic analysis from the response functions \cite{marques2021}. The latter approach can be tiring and slow, needing an extensive calculation of physical variables to be analysed. Moreover, different variables have to be calculated and analysed for distinct transitions. Nevertheless, the NN approach presents itself as a faster alternative, requiring always the same set of parameters to identify all the phases the systems can assume. 

Additionally, since the model predicts phases of individual particles within a system, a population analysis can be performed, from which we showed it is possible to discern different kind of transitions (discontinuous or continuous transitions) and for the region where high and low density liquids appear, if a transition is taking place or the Widom Line is crossed.   

The implementation applied is complementary to works that use a machine learning approach to study water in the supercooled regime, such as those in references \cite{fulford2019, martelli2020}, and explore a new applicability of the binary-mixture network developed in 
\cite{boattini2018}. 

%

\section*{Authors Contribution}

VFH worked on the the conceptualization, methodology, programming and software development, data acquisition and analysis, validation, writing of the original draft, revision and editing. MSM contributed with data acquisition and analysis, writing review and editing. JRB worked on the conceptualization, methodology, programming and software, data acquisition and analysis, writing review and editing, supervision, funding acquisition and project administration.

\section*{Acknowledgments}
Without the public funding this research would not be impossible. VFH thanks the Coordena\c c\~ao de Aperfei\c coamento de Pessoal de N\'ivel Superior (CAPES), Finance Code 001, for the MSc Scholarship. JRB acknowledge to Conselho Nacional de Desenvolvimento Cient\'ifico e Tecnol\'ogico (CNPq) and Funda\c c\~ao de Apoio a Pesquisa do Rio Grande do Sul (FAPERGS) for financial support. All simulations were performed in the SATOLEP Cluster from the Group of Theory and Simulation in Complex Systems from UFPel.

\section*{Data Availability Statement}

All data and codes used in this work are available upon reasonable request.

\section*{References}

\bibliographystyle{unsrt}
\bibliography{refs}

\end{document}